# Gravitational wave stochastic background from cosmological particle decay

Bruce Allen *
*Max Planck Institute for Gravitational Physics, Callinstrasse 38, 30167 Hannover, Germany*

We assume that the cosmological dark matter is composed of massive neutral scalar particles that decay into two massless particles. The decay produces a stochastic background of gravitational waves (GWs) via a "memory effect" mechanism. We calculate the spectral amplitude and slope of the resulting background, which is frequency independent (flat). We discuss its potential observability and show that the resulting background might dominate the cosmological GW background at frequencies above $\approx 10^{10}$ Hz. Penrose has proposed a cosmological model in which dark matter particles have the Planck mass and decay into two gravitons. For these, the spectrum has an additional "direct" contribution from the decay products, which we also calculate. At low frequencies, this direct contribution also has a flat spectrum but with a much smaller amplitude than the memory part.

## I. INTRODUCTION

There is strong evidence that approximately 25% of the cosmological fluid is composed of "dark matter" [1]. This behaves as pressureless dust with a mass density

$$\rho_{\rm DM} \approx 2.1 \times 10^{-30} \text{ g/cm}^3,$$

but its composition is unknown.

This Rapid Communication was originally motivated by Penrose's "conformal cyclic cosmology" [2], which suggests that dark matter is composed of "erebons." These have Planck mass

$$M_{\rm P} = \sqrt{\frac{\hbar c}{G}} = 2.2 \times 10^{-5} \text{ g},$$

and only interact gravitationally. Erebons are not stable, but have a lifetime $\tau$ which is longer than the Hubble time $T_{\rm H} = H^{-1} = 4.6 \times 10^{17}$ s.

Since erebons decay into gravitons, they should leave behind a stochastic cosmological background of gravitational waves (GWs). Penrose conjectured that these might produce a detectable background in the Laser Interferometer Gravitational-Wave Observatory (LIGO) GW detectors. In Sec. III, we calculate the spectrum of that GW background, and show that it is far too weak to observe with LIGO or other instruments.

A surprising finding is that even if dark matter decays into massless particles that are *not* gravitons (for example, photons), a GW background is produced by an effect called "GW memory." In Sec. IV, we calculate that contribution to the GW spectrum, which dominates the previous part at the low frequencies accessible to current GW detectors. Because of the very quick transition associated with particle decay, this process produces a spectrum which extends up to frequencies which are much higher than other known astrophysical sources of GW stochastic background. Our results apply to any neutral scalar dark matter particle which decays into two massless particles.

In this Rapid Communication, $c$ denotes the speed of light, $G$ is Newton's gravitational constant, and $\hbar$ is the reduced Planck's constant.

## II. SPECTRAL FUNCTION

In the literature, the stochastic background of GWs is often characterized with a dimensionless spectral function $\Omega_{\rm GW}(f)$, which describes how the GW energy is distributed in frequency [3,4]. If $d\rho_{\rm GW} c^2$ is the energy density in GWs in the frequency interval $[f, f + df]$, then

$$\Omega_{\rm GW}(f) = \frac{f}{\rho_{\rm c}} \frac{d\rho_{\rm GW}}{df}, \qquad (1)$$

where $\rho_{\rm c} = 3H^2/8\pi G = 8.6 \times 10^{-30}$ g/cm$^3$ is the closure/critical density for which the universe is spatially flat. We use the Planck satellite value $H = 67.4$ km s$^{-1}$ Mpc$^{-1}$ [1] for the current Hubble expansion rate.

Current observations with advanced LIGO, combining O1 and O2 data, constrain $\Omega_{\rm GW}(f)$ for frequencies $f$ below a few hundred Hz [5]. For us, the most relevant limit comes from Fig. 2 and Table 3 of Ref. [5] (with slope $\alpha = 1$). For the spectrum $\Omega_{\rm GW}(f) = (f/25 \text{ Hz})\Omega_0$ the 95% uniform-prior confidence limit is $\Omega_0 < 4 \times 10^{-8}$. This can be directly compared with model predictions.

There are additional low-frequency limits from pulsar timing arrays [6], and broadband limits from big bang nucleosynthesis modeling [7].

---

## III. THE "DIRECT" SPECTRUM

We assume that the decay of the dark matter particle of mass $m$ results in two massless particles with equal and opposite momenta. If those massless particles are gravitons, then these gravitons themselves provide a spectrum of GWs. Here, we calculate that direct graviton spectrum. For particles which decay into other massless particles (for example, photons), please skip to Sec. IV.

In the rest frame of the dark matter particle, the two gravitons resulting from the decay have equal energy and opposite linear momentum. The mean frequency of the two particles $\bar{f}$ follows from energy conservation:

$$\bar{f} = \frac{mc^2}{4\pi\hbar} = 1.5 \times 10^{42} \left(\frac{m}{M_{\rm P}}\right) \text{ Hz}.$$

In the rest frame of the decaying particle, the gravitons have a frequency close to $\bar{f}$ and an energy close to $mc^2/2$. The distribution around these values has a width determined by the particle lifetime $\tau$.

Since we are assuming that these particles compose the dark matter, which has survived to the present, the lifetime $\tau$ must be greater than the Hubble time $T_{\rm H}$. This means that the width $\Delta f = 1/2\pi\tau$ of the emission line is very narrow, $\Delta f \ll \bar{f}$. Nevertheless there is some small probability that the gravitons will be produced at a frequency low enough to detect with LIGO.

In this narrow-line case, the energy distribution of gravitons is described by the Lorentzian distribution

$$L(x) = 1/\pi(1+x^2), \quad (2)$$

where $x = 2(f - \bar{f})/\Delta f$. The expected energy in the frequency range $[f, f + df]$ is $4\pi\hbar\bar{f}L(x)dx$, where $dx = 2df/\Delta f$. Thus, in a time interval $dt$, the contribution to the GW stochastic background energy density from decaying particles in a frequency band $[f, f + df]$ is

$$c^2 d\rho_{\rm GW} = 4\pi\hbar\bar{f} L\left(\frac{2(f - \bar{f})}{\Delta f}\right) \frac{2df}{\Delta f} n \frac{dt}{\tau}, \quad (3)$$

where $n = \rho_{\rm DM}/m$ is the number of particles per unit volume.

To integrate this over time, we need to account for cosmological expansion and redshift. Describe the universe with a spatially flat Friedmann-Robertson-Walker metric $ds^2 = -dt^2 + a^2(t)\vec{dx}^2$, where $t$ is proper time measured by an observer comoving with the cosmological fluid. Let $1 + z = a(t)/a(t')$, where $z$ denotes the redshift of an event at time $t'$ where $t$ is the present time.

Now consider how a contribution to the graviton energy density at time $t'$ in the past affects the current value. Because gravitons are massless, their energy-density redshifts as the fourth power of the scale factor, so $d\rho_{\rm GW} = (1+z)^{-4}d\rho'_{\rm GW}$. The graviton energy density in the past $d\rho'_{\rm GW}$ is obtained by evaluating the right-hand side of Eq. (3) in the past. But since the number density of dark matter redshifts as dust, $n' = (1+z)^3 n$, and since the energy in a given frequency range $[4\pi\hbar\bar{f}L(x)dx]'$ scales as $(1+z)$, all the redshift factors cancel. So one obtains the total energy density today by integrating $dt$ using the current value for $n$.

Integrating the energy density over the Hubble time, and expressing it in terms of the dimensionless spectral function Eq. (1), yields

$$\Omega_{\rm GW}(f) = 2\frac{f}{\Delta f}\left(\frac{\rho_{\rm DM}}{\rho_{\rm c}}\right)\left(\frac{T_{\rm H}}{\tau}\right) L\left(\frac{2(f - \bar{f})}{\Delta f}\right). \quad (4)$$

By integrating over frequency, it can be easily verified that the total energy density in GWs is

$$c^2 \int d\rho_{\rm GW} = c^2 \rho_{\rm c} \int \frac{1}{f}\Omega_{\rm GW} df = \rho_{\rm DM} c^2 \frac{T_{\rm H}}{\tau}.$$

This is what we expect from energy conservation, since in this model the energy in GWs comes from the decay of the dark matter; in the Hubble time the fraction that has decayed is $T_{\rm H}/\tau$.

The predicted $\Omega_{\rm GW}(f)$ at low frequency can be compared with LIGO sensitivity. At low frequencies one finds $L(x) \approx 1/\pi x^2 \approx \Delta f^2/4\pi\bar{f}^2$, so for $f \ll \bar{f}$ one has

$$\Omega_{\rm GW}(f) = \frac{1}{4\pi^2}\frac{f}{\tau\bar{f}^2}\left(\frac{\rho_{\rm DM}}{\rho_{\rm c}}\right)\left(\frac{T_{\rm H}}{\tau}\right). \quad (5)$$

For the erebon, since the lifetime $\tau$ must be greater than the Hubble time, we find $\Omega_{\rm GW}(25 \text{ Hz}) < 2 \times 10^{-103}$. Comparing with the current observational limit in Sec. II shows that this is below detectability in LIGO by about 95 orders of magnitude.

## IV. GW "MEMORY" EFFECT

There is another source of GWs which puts more of the energy at low frequencies. The decay of a massive particle into two massless particles produces a sudden change in the gravitational field, which propagates outwards at the speed of light. Similar effects were first described by Zel'dovich and Polnarev [8] in the slow motion approximation, in the context of massive sources such as supernovae. Braginskii and Thorne [9] called such sources "GW bursts with memory." Later work by Christodoulou [10] showed that there was an additional effect related to the loss of energy by the source.

For massive particle decays into two point particles, one of which is massless, the gravitational wave form was calculated exactly in the weak-field limit by Tolish, Bieri, Garfinkle, and Wald [11], building on earlier work by Tolish and Wald [12]. (While Ref. [11] considers massive particle decay into one massive and one massless particle, the limit as both decay products become massless is well defined and sensible.)

If the initial (massive) particle is at rest at distance $r$ from it, an interferometric GW detector registers a step function (filtered through the response of the detector) at time $r/c$ after the decay. The magnitude of the step is easily determined. Place the origin of coordinates at the massive particle and align the $z$ axis along the path of the massless decay products. Then Eq. (12) of Ref. [11] (set $E = mc^2/2$, since both decay products are massless, and $c = 1$, since that is the convention of Ref. [11]) gives the displacement $\Delta L^a$ in a detector arm $L^a$, located at distance $r$ from the particle, as

$$\Delta L^a = \frac{Gm}{rc^2}(\theta^a\theta_b - \phi^a\phi_b)L^b, \quad (6)$$

where $\theta^a$ and $\phi^a$ are orthonormal spatial vectors tangent to the sphere of radius $r$ at the location of the detector, as illustrated in Fig. 1.

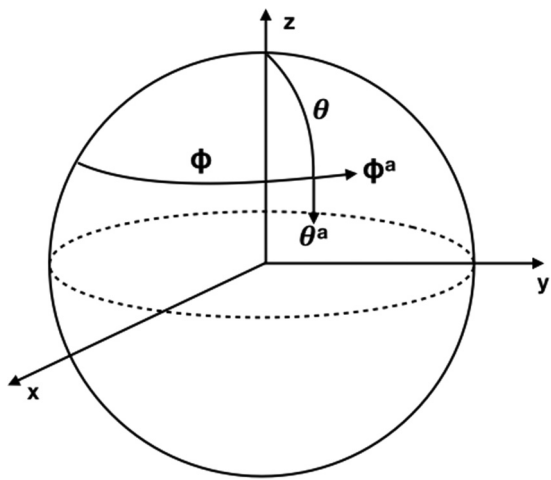

FIG. 1. The unit vectors $\theta^a$ and $\phi^a$ are orthogonal spatial vectors tangent to the sphere at the point $(\theta, \phi)$. $\theta^a$ points along the line longitude $= \phi$ and $\phi^a$ points along a line latitude $= \theta$.

Detector arms based on optical delay lines or Fabry-Pérot cavities sense length changes in the parallel directions. So contracting both sides of this equation with $L^a$ gives the observed GW strain $\Delta L/L$ in one arm as

$$\frac{\Delta L}{L} = \frac{Gm}{rc^2}[(\hat{L}^a\theta_a)^2 - (\hat{L}^a\phi_a)^2],$$

where $\hat{L}^a$ is a unit-length vector along the arm. A detector arm which points in the radial direction is unaffected, a detector arm which points along the longitudinal direction gets longer, and a detector arm which points along a line of latitude gets shorter.

In a LIGO-like detector with two equal-length perpendicular arms $M$ and $N$, the differential strain is given by

$$h = \frac{\Delta M}{M} - \frac{\Delta N}{N}$$
$$= \frac{Gm}{rc^2}[(\hat{M}^a\theta_a)^2 - (\hat{N}^a\theta_a)^2 - (\hat{M}^a\phi_a)^2 + (\hat{N}^a\phi_a)^2].$$

Squaring this quantity and averaging over detector orientations gives a mean-squared strain $\langle h^2 \rangle = 4G^2m^2/5r^2c^4$. This is equivalent to a root-mean-square (rms) strain

$$h_{\rm rms} = \frac{2}{\sqrt{5}}\frac{Gm}{rc^2}. \quad (7)$$

This is easy to check. Although the source is axisymmetric, the average effect on a GW detector (after rotation through all possible orientations of the detector) is independent of $\theta$ and $\phi$ and only depends upon $r$. This follows because Eq. (6) has no $\theta$ or $\phi$ dependence. Examination of Eq. (33) of Ref. [11] shows that the angular dependence of the two null particles $1 \pm \cos\theta$ cancels.

### A. GW "memory" spectrum

If the dark matter is composed of particles which decay in this way, then they will act as a GW source and give rise to a uniform and isotropic background of stochastic GWs. To find its spectrum, we calculate the autocorrelation function $C(\mathcal{T}) = \langle h(t)h(t + \mathcal{T})\rangle$ of the strain $h(t)$ in one detector.

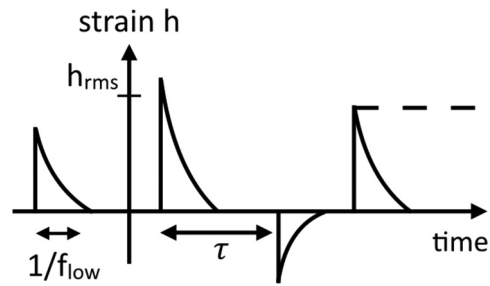

FIG. 2. The "memory effect" wave form in a GW detector from a set of particles decaying at fixed distance $r$ with mean time $\tau$ between events. The rms amplitude is given in Eq. (7); the effective duration of the decaying exponential spikes is the inverse of the detector's low-frequency cutoff $f_{\rm low}$. The signal strength and sign depend upon the detector orientation with respect to the decay. The physical signal consists of the sum of these events and similar ones with lower amplitude and higher rate, coming from larger distances. The dotted line for the final spike shows the step function that would result if the low-frequency cutoff were set to zero.

We begin by looking at the strain signal that would arise from dark matter particles decaying at a fixed distance $r$ from the detector, with mean time $\tau$ between events. We model the detector response as a passive high-pass (one pole, first order) filter, with a low-frequency cutoff $f_{\rm low} = 20$ Hz. The response to a unit gravitational wave step function is a decaying exponential $\exp(-2\pi t f_{\rm low})$.

The detector's strain response $h(t)$ is a series of exponentially decaying spikes as illustrated in Fig. 2. Overlapping this wave form with a time-delayed copy and averaging gives the autocorrelation for sources at radius $r$:

$$C_r(t) = h_{\rm rms}^2 \exp(-2\pi|t|f_{\rm low})/4\pi\tau f_{\rm low}.$$

Fourier transforming this into the frequency domain yields

$$\tilde{C}_r(f) = \int_{-\infty}^{\infty} C_r(t)e^{2\pi i f t}dt = h_{\rm rms}^2/4\pi^2\tau(f^2 + f_{\rm low}^2).$$

For an ideal detector ($f_{\rm low} \to 0$) one obtains

$$\tilde{C}_r(f) = \frac{h_{\rm rms}^2}{4\pi^2\tau f^2} = \frac{G^2m^2}{5\pi^2 r^2 c^4 \tau f^2} \quad (8)$$

for the autocorrelation function of sources at distance $r$. (Note that any detector model which approaches a unity-gain filter as $f_{\rm low} \to 0$ would give the same result.)

Since the sources are incoherent, we can sum their contributions out to the Hubble radius $cT_{\rm H}$ to get the frequency-domain autocorrelation function of the detector strain:

$$\tilde{C}(f) = \int_0^{cT_{\rm H}} \tilde{C}_r(f) n \, 4\pi r^2 dr = \frac{4G^2 m^2 n T_{\rm H}}{5\pi\tau c^3 f^2}, \quad (9)$$

where $n = \rho_{\rm DM}/m$ is the current number of sources per unit volume. This correctly accounts for cosmological expansion and redshift in spatially flat cosmological models, following the same arguments given after Eq. (3).

The frequency-domain autocorrelation function of the strain in one detector is related to the spectral function

$\Omega_{\rm GW}(f)$ using Eq. (3.59) of Ref. [4], with overlap reduction function $\gamma = 1$:

$$\Omega_{\rm GW}(f) = \frac{20\pi^2}{3H^2} f^3 \tilde{C}(f) = \frac{1}{\pi} \frac{\rho_{\rm DM}}{\rho_{\rm c}} \frac{T_{\rm H}}{\tau} \frac{m}{M_{\rm P}} \frac{f}{f_{\rm P}}, \quad (10)$$

where we have defined the Planck frequency as $f_{\rm P} = \sqrt{\frac{c^5}{\hbar G}}/2\pi = 3.0 \times 10^{42}$ Hz. This spectrum is "white": It describes a flat (uniform) distribution of energy as a function of frequency. In Sec. IV B we discuss the high-frequency cutoff of this expression.

For a Planck-mass erebon $m = M_{\rm P}$, at LIGO frequencies, this "indirect" contribution to the GW stochastic background is 60 orders of magnitude larger than the "direct" contribution given in Eq. (5). Unfortunately it is still very small. In comparison with the current LIGO limits in Sec. II, $\Omega_{\rm GW}(25~{\rm Hz}) < 6.7 \times 10^{-43}$ misses detectability by 35 orders of magnitude.

Reference [2] argues that the GW spectrum for particle decay can be found using the "impulsive Vaidya metric" [13], in which a spherically symmetric Schwarzschild spacetime converts itself to Minkowski spacetime along an outgoing spherically symmetric null hypersurface. Such a spherically symmetric particle decay model differs from the axisymmetric model presented here. While we agree that a GW detector would respond to such null dust, Eq. (2.43c) of Ref. [14] shows that for an isotropic source there is no $O(1/r)$ GW term in the strain. The leading effect in strain is $O(1/r^2)$, and thus is not a GW and is negligible for distant sources. Here, we obtain a memory effect specifically because the two outgoing null particles define a preferred axis: The spacetime has axial rather than spherical symmetry.

### B. Total energy converted to GWs

The GW energy spectrum we have found for a single idealized decay (and for the resulting stochastic background) is white, meaning that the energy $c^2 d\rho_{\rm GW}$ in a frequency band $df$ is independent of frequency $f$. Physically, the spectrum is cut off at high frequency because the processes associated with the decay have a shortest characteristic timescale. Here, we examine this more closely.

It follows from Eq. (6) that the transverse-traceless metric perturbation takes the form

$$h_{ab} = \frac{2Gm}{rc^2}(\theta_a\theta_b - \phi_a\phi_b)W(t - r/c), \quad (11)$$

where in the rest frame of the decaying particle $t$ is time after the decay, $r$ is radial distance from the decay, and $W(x)$ is the "profile" of the expanding circular wave front: a smoothed dimensionless step function which vanishes for negative argument and approaches unity for large positive argument.

In this gauge the energy density in GWs is $c^2 \dot{h}_{ab}\dot{h}^{ab}/32\pi G$, where $\dot{} = d/dt$. Integrate this over a spatial slice at time $t > 0$, assuming that $t$ is large enough to ensure that the outgoing radiation has formed a shell away from the origin. One obtains a total energy

$$E_{\rm GW} = \frac{Gm^2}{c} \int_0^\infty [\dot{W}(t)]^2 dt, \quad (12)$$

where $\dot{W}(t) = dW(t)/dt$.

To evaluate this integral we use a simple linear model wave form $W(t)$ which enforces causality [15], so vanishes for $t < 0$:

$$W(t) = \begin{cases} 0 & \text{for } t < 0, \\ t/\Delta t & \text{for } 0 \leqslant t < \Delta t, \text{ and} \\ 1 & \text{for } \Delta t \leqslant t. \end{cases} \quad (13)$$

Here, $\Delta t$ is the time duration of the outgoing pulse for an observer at fixed radius $r$.

The integral over time that appears in Eq. (12) yields $1/\Delta t$, so the outgoing GW carries energy $E_{\rm GW} = Gm^2/c\Delta t$. If the time duration $\Delta t$ of the pulse is related to the rest-mass energy $mc^2$ of the decaying particle via the Heisenberg uncertainty principle, then $\Delta t \geqslant \hbar/mc^2$ and the energy carried away is bounded by

$$E_{\rm GW} \leqslant \frac{c\,Gm^3}{\hbar} = \left(\frac{m}{M_{\rm P}}\right)^2 mc^2. \quad (14)$$

Note that for particles of mass smaller than Planck mass, GWs carry away only a small fraction of the total energy, but for particles close to the Planck mass, a significant fraction of the energy could be lost to GWs.

## V. CONCLUSIONS

We have calculated the spectrum of GWs produced by the decay of massive dark matter particles into two massless ones. Because of the "memory effect," the decay produces an "indirect" stochastic background of GWs with a flat spectrum, and spectral function $\Omega_{\rm GW}(f)$ given in Eq. (10). If the massless particles are gravitons, then the decay also generates a much smaller "direct" GW spectrum, given in Eq. (5). This is also flat if the gravitons produced by the decay have frequencies which are high compared to the detection band. For massive particles, the indirect part of the spectrum dominates.

Because of the "sharpness in time" of the transition, a "memory effect" background (but with reduced amplitude) should also be produced by a decay into massive rather than massless particles.

The "memory effect" background is important because standard astrophysical sources do not produce high frequency GWs, and neither do conventional early-universe models. For example, the GW background produced in slow-roll inflationary models only extends to $\approx$GHz frequencies (see the final two paragraphs of Sec. II of Ref. [16]). In cosmic string models and defect models, the GW background falls off as a power law at high frequencies [17].

So, although it is weak, the background from particle decay may nevertheless dominate the high-frequency GW spectrum. While current technology offers no prospects for constraining or detecting such backgrounds, in the future its high-frequency nature might make it accessible to new experimental methods and techniques.

### ACKNOWLEDGMENTS

We thank Thibault Damour, David Garfinkle, and Bob Wald for helpful conversations and corrections, and Anna Ijjas

and Paul Steinhardt for an invitation to the Princeton/Simons meetings which motivated this work. We also thank the anonymous referee for identifying several mistakes in the original version of this manuscript.